\documentclass[pra,aps,amsfonts,amsmath,superscriptaddress,%
  twocolumn,showpacs,floatfix]{revtex4}
\usepackage{amsfonts}
\usepackage{amsmath}
\usepackage{bm}
\usepackage{epsfig}
\newcommand{\vek}[1]{\bm{\mathrm{#1}}}
\begin{document}
\title{Non-standard pairing in asymmetric trapped Fermi gases}
\author{Paolo Castorina}
\affiliation{Dipartimento di Fisica e Astronomia, Via Santa
  Sofia 64, I-95123 Catania, Italy}
\author{Marcella Grasso}
\affiliation{Dipartimento di Fisica e Astronomia, Via Santa
  Sofia 64, I-95123 Catania, Italy}
\affiliation{INFN, Sezione di Catania, Via Santa
  Sofia 64, I-95123 Catania, Italy}
\affiliation{Institut de Physique Nucl\'eaire, 15 rue Georges
  Cl\'emenceau, F-91406 Orsay Cedex, France}
\author{Micaela Oertel}
\affiliation{Service de Physique Nucl\'eaire,
CEA-DAM Ile-de-France, BP 12, F-91680 Bruy\`eres-le-Ch\^atel, France}
\author{Michael Urban}
\affiliation{Institut de Physique Nucl\'eaire, 15 rue Georges
  Cl\'emenceau, F-91406 Orsay Cedex, France}
\author{Dario Zappal\`a}
\affiliation{INFN, Sezione di Catania, Via Santa
  Sofia 64, I-95123 Catania, Italy}
\begin{abstract}
We study an ultracold trapped Fermi gas of atoms in two hyperfine
states with unequal populations. In this situation the usual BCS
pairing is suppressed and non-standard pairing mechanisms become
important. These are treated by solving the Bogoliubov-de Gennes
equations, which at the same time correctly take into account the
finite size of the trapped system. We find results which can be viewed
as generalization of the LOFF phase to finite systems.
\end{abstract}
\pacs{03.75.Ss,21.60.Jz}
\maketitle

In the last years much progress has been made in improving the
techniques used to trap and cool dilute gases of bosonic and fermionic
atoms \cite{1999,Bmodes}. One of the interesting aspects of the
properties of ultra-cold gases is that the interatomic interaction can
be modified, both in its intensity and in its sign, by changing the
applied magnetic field around a Feshbach resonance. Due to the very
low densities and temperatures in these systems the details of the
interatomic interaction are unimportant and the interaction can be
characterized by one single parameter, the $s$-wave scattering length
$a$. In this article we consider Fermi gases trapped and cooled in two
hyperfine states with an attractive interaction, i.e., $a < 0$. We
will concentrate on the weakly interacting case ($k_F |a| \ll 1$,
where $k_F$ denotes the Fermi momentum). In this region, BCS
superfluidity with formation of Cooper pairs is expected below a
certain critical temperature. So far, some experimental signals have
been found which would indicate the existence of superfluidity in this
region \cite{Bmodes}, but a clear evidence is still missing.

Besides the interaction, also the population of the two hyperfine
states can experimentally be controlled. Usually mixtures with equal
populations are created, but controlled asymmetric mixtures have also
been obtained \cite{zw}. Unequal populations of the two hyperfine
states can lead to very interesting phenomena. For instance, the BCS
pairing mechanism is supposed to become suppressed \cite{houbiers}
since the two Fermi momenta associated with the two spin polarizations
become different: The formation of zero-momentum Cooper pairs built of
two atoms at their respective Fermi surface becomes difficult.
Instead, other more exotic pairing phenomena have been suggested for
the case of unequal populations, like the
Larkin-Ovchinnikov-Fulde-Ferrel (LOFF) phase \cite{combescot}, the
Sarma (interior gap) phase \cite{liu,intgap}, or a phase with deformed
Fermi surfaces (DFS) \cite{dfs}. Many of these non-standard pairing
mechanisms have already been discussed in other domains of physics
where asymmetric two-component fermion systems can be found:
Superconductors in a magnetic field \cite{loff,sarma}, neutron-proton
pairing in asymmetric nuclear matter \cite{selo}, color
superconductivity in high density QCD \cite{csc}. The experimental
observation of the LOFF phase in the case of superconductivity is a
subject of debate. It seems that quite recently an oscillating
superconducting order parameter has been observed in a ferromagnetic
thin film \cite{kontos}.

Usually~\cite{houbiers,combescot,liu,intgap,dfs}
these non-standard pairing types in ultracold Fermi gases are
usually discussed for the case of infinite and homogeneous systems,
or for trapped systems in local-density approximation (LDA) which
again amounts to treating the system locally as homogeneous. However,
as we are going to show, in certain cases the LDA can become a very
poor approximation and we therefore want to analyze this problem
within a fully microscopic mean field Bogoliubov-de Gennes (BdG)
approach \cite{deGennes} taking into account the inhomogeneity and
finite size of the trapped system. Recently the solution of the BdG
equations has been considered in Ref.~\cite{mizushima}, where the authors 
discuss also possibilities for the experimental detection of a spatially
modulated pairing gap. 

In the present article we study two systems: a small one where shell
effects still play a role and a relatively large one, where one could
expect the LDA to be a reasonable approximation. As we will show, this
is not always the case, although the LDA describes roughly
the qualitative features. In addition, we examine the temperature
dependence of the non-standard pairing phase, since this is important
in connection with the experimental realization of such a phase.

Let us briefly recall the BdG formalism. We consider a system
containing $N = N_+ + N_-$ atoms of mass $m$ trapped by a spherical
harmonic potential in two hyperfine states denoted by $+$ and $-$. The
many-body Hamiltonian of the system can be written as
\begin{multline}
H = \sum_\sigma\int d^3 r \Big(\Psi^\dagger_\sigma(\vek{r}) H_0
  \Psi_{\sigma} (\vek{r})\\
+ g\Psi^\dagger_+ (\vek{r}) \Psi^\dagger_- (\vek{r})
  \Psi_- (\vek{r})\Psi_+ (\vek{r})\Big)~,
\label{H}
\end{multline}
where $H_0=-\hbar^2\nabla^2/(2m) + m\omega^2 r^2/2$ denotes the
harmonic oscillator single-particle Hamiltonian,
$\Psi_{\sigma}(\vek{r})$ is the field operator which annihilates a
particle at the position $\vek{r}$ in the spin state $\sigma$ ($+$ or
$-$), $g=4\pi\hbar^2 a/m$ is the coupling constant. In mean-field
approximation, on can derive the following BdG equations corresponding
to the Hamiltonian (\ref{H}):
\begin{equation}
\begin{split}
u_{\eta\sigma} (\vek{r}) E_{\eta \sigma}&=W_\sigma u_{\eta
  \sigma}(\vek{r}) +\Delta(\vek{r}) v_{\eta-\sigma}(\vek{r})~,\\
  v_{\eta-\sigma}(\vek{r}) E_{\eta \sigma}&=-W_{-\sigma}
  v_{\eta-\sigma}(\vek{r}) +\Delta(\vek{r}) u_{\eta \sigma}(\vek{r})~,
\label{bdg}
\end{split}
\end{equation}
where $W_\sigma = H_0 + g \rho_{-\sigma}(\vek{r}) - \mu_\sigma$,
$\mu_\sigma$ and $\rho_\sigma$ being the chemical potential and the
density, respectively. $\Delta(\vek{r})$ denotes the pairing field
(gap) and $E_{\eta\sigma}$, $u_{\eta\sigma}$ and $v_{\eta\sigma}$ are
the quasiparticle energy and wave functions, respectively,
corresponding to the quantum numbers $\eta = n,l,m$ and spin
$\sigma$. In order to have different populations, the two chemical
potentials $\mu_+$ and $\mu_-$ must be different. It is convenient to
introduce the average chemical potential $\bar{\mu}$ and to write
$\mu_\sigma = \bar{\mu}+\sigma \delta \mu/2$
where $\delta\mu$ determines the asymmetry. Eqs. (\ref{bdg}) reduce to
the usual BdG equations in the symmetric case $\mu_+ = \mu_-$. They
are solved numerically employing the same regularization method for
the pairing field as described in Ref.~\cite{grur} for the symmetric
case.

Eqs. (\ref{bdg}) are general enough to describe rather complicated
types of non-standard pairing. In the case of usual BCS pairing, the
dominant matrix elements of the pairing field are the diagonal ones,
i.e., each state $|n,l,m,+\rangle$ is paired with its time-reversed
counterpart $|n,l,-m,-\rangle$. However, the non-diagonal matrix
elements of $\Delta$ are also included, which amounts to taking into
account also the pairing of states $|n,l,m,+\rangle$ and
$|n',l,-m,-\rangle$ with different principal quantum numbers $n'\neq
n$. In our present calculation, we still keep the restriction that the
Cooper pairs have total angular momentum zero. To release this
constraint would mean to allow for a spontaneous breakdown of
spherical symmetry, which would be numerically very heavy. The effect
of strong non-diagonal matrix elements of $\Delta$ in fact corresponds
closely to the LOFF phase in the case of a uniform system. There, the
states are labeled by their momentum $\vek{k}$. In the simplest
version of the LOFF phase, the Cooper pairs have total momentum
$\vek{q}$, i.e., each state $|\vek{k},+\rangle$ is paired with
$|-\vek{k}+\vek{q},-\rangle$. The corresponding gap is oscillating
with wave vector $\vek{q}$ and its matrix elements are therefore of
the form $\Delta_{\vek{k}\vek{k}'} = \Delta
\delta_{\vek{k}-\vek{q},\vek{k}'}$ (in contrast to the BCS phase,
where $\Delta_{\vek{k}\vek{k}'} = \Delta\delta_{\vek{k}\vek{k}'}$)
\footnote{Note, however, that this analogy between the trapped and the
homogeneous system is not perfect, since in the trapped system with
$\delta \mu \neq 0$ even pairing between states $|n,l,m,+\rangle$ and
$|n,l,-m,-\rangle$ includes pairing of states with different wave
functions due to the different mean fields felt by atoms with
different spin projections.}.

In the discussion of our results all quantities will be given in
harmonic oscillator units. We use the same coupling constant as in
Ref. \cite{grur}, i.e., $g = -\hbar^2 l_{ho}/m$, where $l_{ho} =
\sqrt{\hbar/(m\omega)}$ denotes the harmonic oscillator length, and we
consider two values for the average chemical potential, $\bar{\mu} =
22$ $\hbar\omega$ (small system with $N \approx 4900$) and $32$
$\hbar\omega$ (large system with $N \approx 17000$).

In Fig. 1 
%%%%%%%%%%%%%%%%%%%%%%%%%%%%%%%%%%%%%%%%%%%%%%%%%%%%%%%%%%%%%%%%%%%%%%%
\begin{figure}
\begin{center}
\epsfig{file=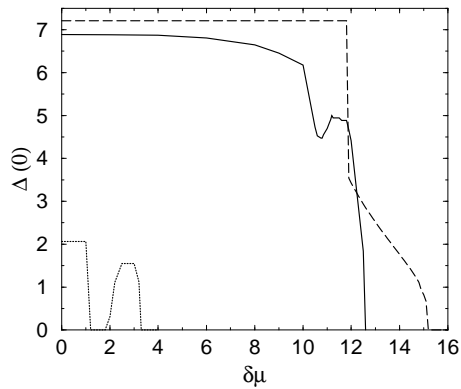,width=6cm}
\end{center}
\caption{Value of the pairing gap at the center of the trap (in units
of $\hbar \omega$) as a function of the asymmetry $\delta \mu$ (in
units of $\hbar \omega$). The lowest line corresponds to $\bar{\mu} =
22\hbar\omega$. The two upper lines correspond to $\bar{\mu} =
32\hbar\omega$ and show the BdG (solid line) and the LDA result
(dashed line).}
\label{fig1}
\end{figure}
%%%%%%%%%%%%%%%%%%%%%%%%%%%%%%%%%%%%%%%%%%%%%%%%%%%%%%%%%%%%%%%%%%%%%%%
we show the values of the pairing gap $\Delta(0)$ at the center of the
trap for increasing aymmetry $\delta \mu$ at $T=0$. Let us first look
at the lowest line, corresponding to the small system with $\bar{\mu}
= 22$ $\hbar\omega$. When both spin states are equally populated
($\delta\mu = 0$), we find $\Delta(0)\approx 2$ $\hbar\omega$, i.e.,
we are no more in the intrashell-pairing regime, but shell effects are
still important \cite{grur}. If we increase $\delta\mu$, the two Fermi
surfaces become more and more separated, i.e., if the state
$|n,l,m,+\rangle$ lies close to the Fermi level for spin $+$, the
state $|n,l,-m,-\rangle$ lies far from the Fermi level for spin $-$,
making BCS pairing less and less favorable. As a consequence, at
$\delta\mu \approx 1.2$ $\hbar\omega$, corresponding to a particle
number asymmetry $\alpha = (N_+ - N_-)/N \approx 0.06$, the pairing
disappears (shell closure effect). But then, near $\delta\mu \approx
2$ $\hbar\omega$ ($\alpha \approx 0.07$), the states $|n,l,m,+\rangle$
near the Fermi level for spin $+$ approach the states
$|n',l,-m,-\rangle$ near the Fermi level for spin $-$ if $n' =
n-1$. Therefore, pairing becomes again possible, but now the Cooper
pairs are built of two wavefunctions with different numbers of nodes,
leading to a gap $\Delta(r)$ which as a function of $r$ has exactly
one node.

Let us now turn to the investigation of the larger system, $\bar{\mu}
= 32$ $\hbar\omega$. Here it seems to be appealing to estimate if and
where the LOFF phase could appear by using the LDA, which should be
exact in an infinite system. In order to do this, we calculate at each
point $R$ the thermodynamic potential $\Omega$ of a uniform gas with
effective average chemical potential $\bar{\mu}_{\mathit{eff}}(R) =
\bar{\mu}-m\omega^2 R^2/2$, assuming a gap of the form
$\Delta(\vek{r}) = \Delta \exp(i\vek{q}\cdot\vek{r})$, and minimize
$\Omega$ with respect to $\Delta$ and $q$. To be more precise, we
should have taken a more sophisticated expression for the order
parameter, but we stress here that we use the LDA just to have some
indications about what kind of behavior one should expect. Let us
again look at $\Delta(0)$ as a function of the asymmetry (dashed line
in Fig. 1). We find that LDA gives the BCS solution $q=0$ as the most
favorable solution up to $\delta \mu =11.9$ $\hbar \omega$. At that
asymmetry we find a first-order phase transition (i.e., a
discontinuity in $\Delta(0)$) to the LOFF phase with $q \sim$
$l_{ho}^{-1}$ which means that the order parameter oscillates with a
wavelength of $\sim 6.2$ $l_{ho}$. This behavior is different from the
microscopic (BdG) result (solid line in Fig. 1), which shows a smooth
behavior of $\Delta(0)$. Nevertheless, also in the BdG calculation
there is a rapid change of $\Delta(0)$ between $\delta\mu = 10$ $\hbar
\omega$ ($\alpha \approx 0.25 $) and $\delta \mu = 11$ $\hbar \omega$
($\alpha \approx 0.29$), where the system goes from the BCS-type to
the LOFF-type phase, as discussed above. The minimum that one observes
for the BdG gap at $\delta \mu \approx 10.8$ $\hbar \omega$ and the
subsequent enhancement are due to shell effects which still persist
even in this large system and which cannot be reproduced by the
semiclassical LDA calculation.

In Fig. 2
%%%%%%%%%%%%%%%%%%%%%%%%%%%%%%%%%%%%%%%%%%%%%%%%%%%%%%%%%%%%%%%%%%%%%%%
\begin{figure}
\begin{center}
\epsfig{file=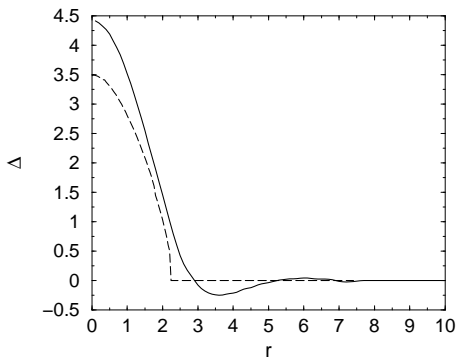,width=6cm}
\end{center}
\caption{Radial profile of the pairing gap $\Delta(r)$ (in units of
$\hbar \omega$) for $\delta \mu =12 $ $\hbar \omega$. The radial
coordinate $r$ is expressed in units of $l_{ho}$. The microscopic
(solid line) and the LDA (dashed line) results are plotted.}
\label{fig2}
\end{figure}
%%%%%%%%%%%%%%%%%%%%%%%%%%%%%%%%%%%%%%%%%%%%%%%%%%%%%%%%%%%%%%%%%%%%%%%
we plot the radial profile of the pairing field $\Delta(r)$ for
$\delta \mu = 12$ $\hbar\omega$, corresponding to $\alpha \approx
0.3$, at $T=0$. The microscopic (solid line) and the LDA (dashed line)
results are shown. Within LDA, in this case, the LOFF phase is more
favorable than BCS for all values of $r$. We observe in Fig. 2 that
the LDA gap goes abruptly to zero at a radius of $\sim 2$ $l_{ho}$,
which is smaller than the LDA wavelength of $\sim 6.2$ $l_{ho}$. Thus,
the region where the gap is non-zero does not even contain one
wavelength of the oscillation and therefore the validity of LDA seems
to be very questionable. As expected from the symmetric case
\cite{grur}, the LDA fails to describe the tail of the pairing field:
The LDA gap goes abruptly to zero while in the microscopic case the
gap has a smooth profile. We finally observe that the microscopic
order parameter makes an oscillation and that a node is situated at
$\approx 3$ $l_{ho}$: The modulation of the order parameter and the
presence of a node are signals which indicate that the system is in a
LOFF-type phase.

Let us consider now the case of a smaller asymmetry, $\delta \mu =6$
$\hbar \omega$, corresponding to $\alpha \approx 0.15$. We show in Fig. 3
%%%%%%%%%%%%%%%%%%%%%%%%%%%%%%%%%%%%%%%%%%%%%%%%%%%%%%%%%%%%%%%%%%%%%%%
\begin{figure}
\begin{center}
\epsfig{file=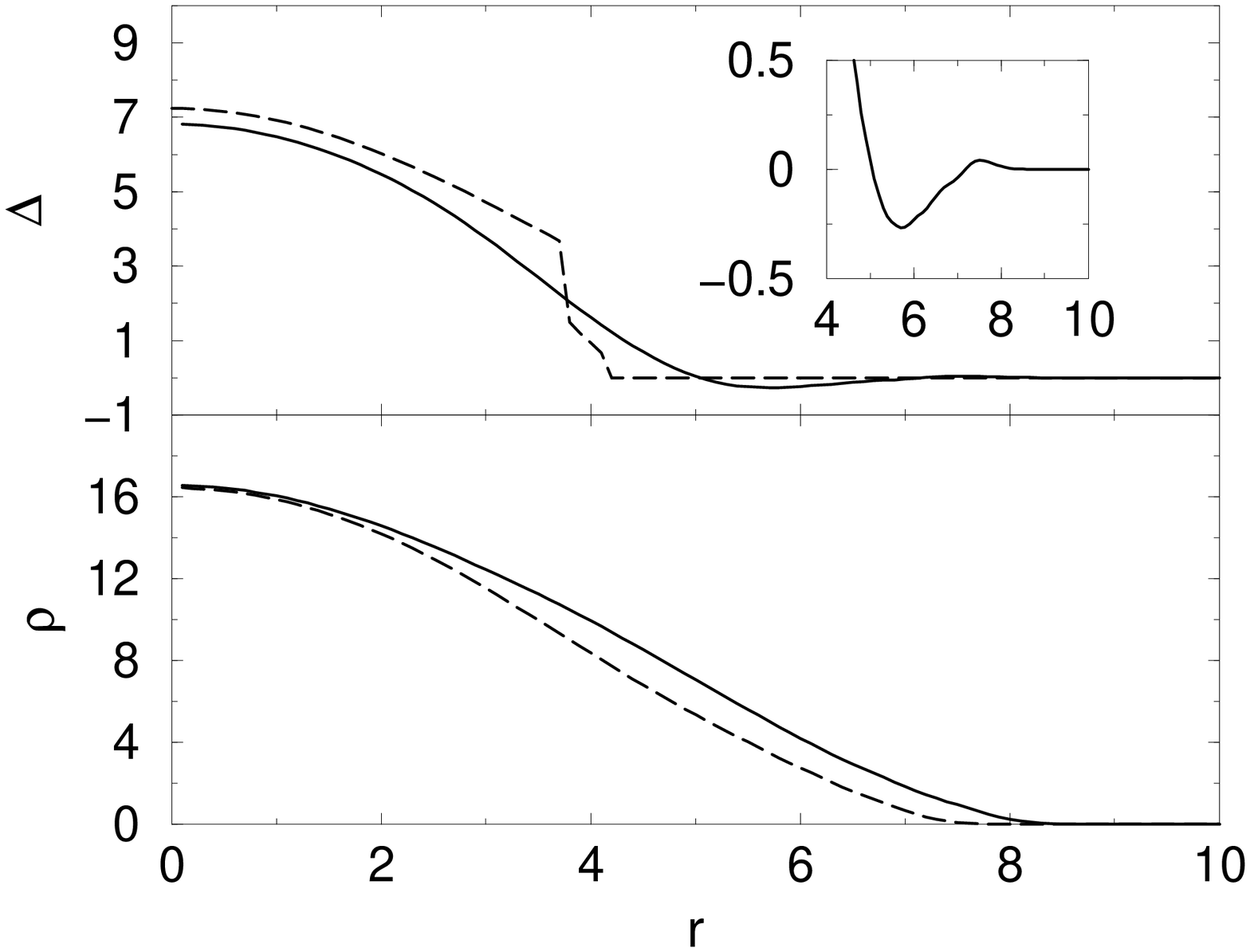,width=6cm}
\end{center}
\caption{Radial profiles of the pairing gap $\Delta(r)$ in units of
$\hbar \omega$ (top) and of the densities in units of $l_{ho}^{-3}$
(bottom) for $\delta \mu = 6$ $ \hbar \omega$ as a function of the
radial coordinate $r$ in units of $l_{ho}$. In the upper panel the
solid and dashed lines correspond to the BdG and LDA results,
respectively. In the lower panel, the solid and the dashed lines refer
to the BdG results for the $+$ and $-$ densities,
respectively.}
\label{fig3}
\end{figure}
%%%%%%%%%%%%%%%%%%%%%%%%%%%%%%%%%%%%%%%%%%%%%%%%%%%%%%%%%%%%%%%%%%%%%%%
the radial profiles of the gap $\Delta(r)$ (top) and of the densities
(bottom) at $T=0$. In the upper panel we report the microscopic (BdG)
gap (solid line) and the LDA result (dashed line). In this case,
according to the LDA, the BCS phase ($q=0$) would be energetically
preferred in the center of the gas (as we have shown in Fig. 1) and up
to $r=3.8$ $l_{ho}$, while the LOFF phase with $q \sim 0.7$
$l_{ho}^{-1}$ would be more favorable in the interval $3.8$ $l_{ho} <
r < 4.1$ $l_{ho}$. For larger values of $r$, the LDA predicts that the
gap should be zero. The wavelength of the oscillation of the order
parameter in the LOFF phase ($\sim 8.9$ $l_{ho}$) would again be much
larger than the region where the gap is non zero. Contrary to the LDA,
the microscopic BdG calculation gives a smooth behavior of the order
parameter. Near the center, it corresponds rather well to the LDA
prediction, indicating that the pairing is more or less of BCS
type. Between $r = 4$ $l_{ho}$ and $10$ $l_{ho}$, the gap is
oscillating (see inset in the upper panel of Fig. 3), indicating the
appearance of the LOFF-type phase. Although within the BdG calculation
there is no sharp transition from one phase to the other,
qualitatively it seems that both phases can be present at the same
time in different regions of the system.

In the lower panel of Fig. 3 the BdG results for the two densities
($\rho_+$ and $\rho_-$) are shown. One observes that in the center of
the gas the two densities are equal. This is coherent with the fact
that in the BCS phase at $T=0$ the LDA always gives $\rho_+ = \rho_-$
if $\Delta > \delta\mu/2$, as it is the case here.

All the results shown so far refer to $T=0$. However, in real
experiments with trapped atomic gases the temperature is always
non-zero. Let us therefore raise the temperature in the case of
asymmetry $\delta \mu = 6$ $\hbar \omega$ in order to analyze what
happens to the gap modulation when the temperature is finite. In
Fig. 4 
%%%%%%%%%%%%%%%%%%%%%%%%%%%%%%%%%%%%%%%%%%%%%%%%%%%%%%%%%%%%%%%%%%%%%%%
\begin{figure}
\begin{center}
\epsfig{file=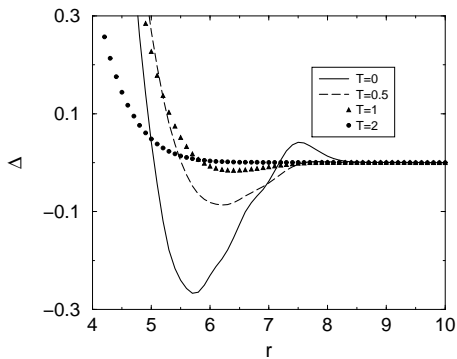,width=6cm}
\end{center}
\caption{Radial profile of the pairing gap $\Delta(r)$ (in units of
$\hbar \omega$) for $\delta \mu = 6$ $ \hbar \omega$. The radial
coordinate $r$ is expressed in units of $l_{ho}$. Results at $T=0$
(solid line), $T=0.5$ $\hbar \omega /k_B$ (dashed line) $T=\hbar
\omega /k_B$ (triangles) and $T=2$ $\hbar \omega /k_B$ (circles) are
reported.}
\label{fig4}
\end{figure}
%%%%%%%%%%%%%%%%%%%%%%%%%%%%%%%%%%%%%%%%%%%%%%%%%%%%%%%%%%%%%%%%%%%%%%%
we show the order parameter in the radial interval from $4$ $l_{ho}$
to $10$ $l_{ho}$ (where we observed an oscillation in the case $T=0$)
for four values of temperature, $T=0$ (solid line), $T= 0.5$ $\hbar
\omega / k_B$ (dashed line) $T= \hbar \omega / k_B$ (triangles) and
$T= 2$ $\hbar \omega / k_B$ (circles). One observes that the
oscillation has a smaller and smaller amplitude with increasing
temperature and that it disappears between $T= \hbar \omega / k_B$ and
$T= 2$ $\hbar \omega / k_B$. Our interpretation of this result is that
the critical temperature of the LOFF-type phase is smaller than the
BCS critical temperature. Therefore, the LOFF-type phase disappears at
some temperature between $ \hbar \omega / k_B$ and 2 $ \hbar \omega /
k_B$, while the BCS gap in the central region of the gas is still
different from zero at $T= 2$ $\hbar \omega / k_B$. In this case the
LDA results (not shown) are very different from the BdG ones (the gap
is much too large), as one could expect from the fact that already in
the symmetric case the agreement between LDA and BdG calculations
becomes quite poor at finite temperature \cite{grur}. Nevertheless,
qualitatively the LDA gives again the right indication: Also within
LDA the LOFF phase is absent at $T=2$ $\hbar \omega / k_B$. Instead,
at that temperature the Sarma phase becomes more favorable in certain
regions within the LDA: This phase is characterized by zero momentum
Cooper pairs, a gap $\Delta$ smaller than $\delta \mu / 2$, different
densities $\rho_+$ and $\rho_-$ and typical occupation number
distributions as shown in Ref. \cite{liu}.

To summarize, we have solved the BdG equations for an atomic Fermi gas
with different populations of two hyperfine states. It is well-known
that an increasing asymmetry of the populations renders BCS pairing
difficult, and non-standard pairing mechanisms become possible. In
this article we showed that the BdG formalism automatically includes
such non-standard pairing mechanisms through the non-diagonal matrix
elements of the gap. For example, in the case of a small system, we
found that the usual pairing disappears at a certain asymmetry, but
when the asymmetry is strong enough such that the single-particle
energies of states with opposite spin and different principal quantum
numbers start to match, pairing becomes again possible, but now with
an oscillating order parameter. This is very similar to the LOFF phase
introduced for the case of a homogeneous system. In the case of a
larger system, there is no longer a sharp separation between the BCS
pairing and the LOFF-type pairing: As a function of asymmetry, but
also as a function of the distance from the center of the trap, the
system undergoes smooth transitions from one kind of pairing to the
other. This result is qualitatively different from that obtained with
LDA calculations, where the transition between the BCS and the LOFF
phase is a first order phase transition. We also observe that even a
system containing 17000 atoms is still much too small for the LDA to
be applicable, since the wavelength of the LOFF oscillations is of the
same order of magnitude as the whole system. Finally we looked at the
temperature dependence of the LOFF-type phase. We observe that it
disappears already at temperatures where the BCS phase is still
present. This, of course, can be a problem if one tries to observe the
LOFF phase in experiments.

We acknowledge discussions with M. Baldo, F. Cataliotti and A. Sedrakian. 


\begin{thebibliography}{*}

\bibitem{1999} B. De Marco and D.S. Jin, Science \textbf{285}, 285
(1999); M. Greiner, C.A. Regal, D.S. Jin, Nature \textbf{426}, 537
(2003); M.W. Zwierlein, et al., Phys. Rev. Lett.\textbf{91}, 250401
(2003); C.A. Regal, M. Greiner, and D.S. Jin,
Phys. Rev. Lett. \textbf{92}, 040403 (2004); M.W. Zwierlein, et al.,
Phys. Rev. Lett. \textbf{92}, 120403 (2004); M. Bartenstein, et al.,
Phys. Rev. Lett. \textbf{92}, 120401 (2004); T. Bourdel, et al.,
Phys. Rev. Lett. \textbf{93}, 050401 (2004).
\bibitem{Bmodes} J. Kinast, S.L. Hemmer, M.E. Gehm, A. Turlapov, and
  J.E. Thomas, Phys. Rev. Lett. \textbf{92}, 150402 (2004);
 C. Chin, et al., Science \textbf{305}, 1128 (2004).
\bibitem{zw} M.W. Zwierlein, et al., Phys. Rev. Lett. \textbf{91}, 
  250404 (2003). 
\bibitem{houbiers} M. Houbiers et al., Phys. Rev. A \textbf{56}, 4864 (1997).
\bibitem{combescot} R. Combescot and C. Mora, Europhys. Lett. \textbf{68},
  79 (2004).
\bibitem{liu} W.V. Liu and F. Wilczek, Phys. Rev. Lett. \textbf{90},
  047002 (2003).
\bibitem{intgap} P.F. Bedaque, H. Caldas, and G. Rupak, Phys. Rev. Lett.
  \textbf{91}, 247002 (2003); J. Carlson and Sanjay Reddy, cond-mat/0503256.
\bibitem{dfs} A. Sedrakian et al., cond-mat/0404577;
\bibitem{loff} P. Fulde and R.A. Ferrel, Phys. Rev.\textbf{135}, A550 (1964);
  A.I. Larkin and Yu.N. Ovchinnikov, Sov. Phys. JETP \textbf{20}, 762 (1965).
\bibitem{sarma} G. Sarma, Phys. Chem. Solids \textbf{24}, 1029 (1963).
\bibitem{selo} A. Sedrakian and U. Lombardo, Phys. Rev. Lett. \textbf{84}, 
  602 (2000).
\bibitem{csc} J.A. Bowers and K. Rajagopal, Phys. Rev. D \textbf{66},
  065002 (2002); I. Shovkovy and M. Huang, Phys. Lett. B \textbf{564},
  205 (2003); R. Casalbuoni and G. Nardulli, Rev. Mod. Phys. \textbf{76},
  263 (2004).
\bibitem{kontos} T. Kontos, M. Aprili, J. Lesueur, and X. Grison, Phys.
  Rev. Lett. \textbf{86}, 304 (2001). 
\bibitem{deGennes} P.-G. de Gennes, \textit{Superconductivity of Metals and 
  Alloys} (Benjamin, New York, 1966).
\bibitem{mizushima} T. Mizushima, K. Machida, and M. Ichioka,
  Phys. Rev. Lett. \textbf{94}, 060404 (2005).
\bibitem{grur} M. Grasso and M. Urban, Phys. Rev. A \textbf{68}, 033610
  (2003).
\end{thebibliography}
\end{document}